Complex Network Analysis of Transportation Networks

# A Review on Transportation Network based on Complex Network Approach

**Nur Umaisara Rashid[1] and Chan Kar Tim[2*]**

[1]Institute for Mathematical Research, Universiti Putra Malaysia, 43400 UPM, Serdang, Selangor, Malaysia
[2]Department of Physics, Faculty of Science, Universiti Putra Malaysia, 43400 UPM, Serdang, Selangor, Malaysia

Email addresses
nur_umaisara_rashid@yahoo.com.my (Nur Umaisara Rashid)
chankt@upm.edu.my (Chan Kar Tim)
*corresponding author

List of Figure:
Figure 1: Examples of transportation network in (a) Brazil by Couto et al. (2015) and (b) China by W.-B. Du et al. (2016)
Figure 2: Examples of a) PTN in 5 different spaces namely b) $l$-space, c) $p$-space, d) $b$-space, e) $c$-space and f) $r$-space
Figure 3: The evolution of Chinese main air transport network (CMATN) in a span of 10 years that demonstrates the increase in passengers volume (Su et al., 2019)

A Review on Transportation Network based on Complex Network Approach

**Abstract**

Complex network theory is being widely used to study many real-life systems. One of the fields that can benefit from complex network theory approach is transportation network. In this paper, we briefly review the complex network theory method assimilated into transportation network research and the analysis it provided. It is irrefutable that complex network theory is capable to explain the structure, dynamic, node significance, performance as well as evolution of the transportation network.

# 1 Introduction

Transportation as a whole is a service that is highly consumed by main cities citizen all over the world. Public transportation especially is an important indicator of the development of a country. While the enhancement of public transportation system could contribute to the country progression, the same way can be said vice versa. In addition, transportation is a very vital aspect of one's lifestyle. Most country have their own systematic public transportation services be it metro, taxi, bus or subway. The existence of a useful public transportation service is key to development of a country as it can help the people who work in the city to commute from their home to their workplace. One of the factors in the need of the emergence of a comprehensive public transportation system that prioritize the transportation service area and the scheduling is the urbanization which encourages by migration. According to (Rashid & Ghani, 2011), that internal migration contributes to the urbanization by decision makers.

Researchers all over the world have been analysing public transportation network based on complex network approach to obtain more information on the services while improving the future planning of the said services (De Bona, Fonseca, Rosa, Lüders, & Delgado, 2016). In this paper, we are focusing on the type of transportation network, and the analysis that have been conducted across the years. Transportation network usually involves a set route or planned schedule that make it easier for the consumers to benefit from the services. Public transportation network has always grabbed the researcher's attention over the past decades, due to their structure that may have expanded in the future according to the local development in the country involved. The current state of the transportation network may also contribute to the better understanding of the whole system while increasing knowledge on complex network theory. This will also help the future planning of the region.

Complex network analysis is one of the most powerful tools used to understand a large range of real complex systems (Newman, 2010). These include the social networks such as friendship network (González, Herrmann, Kertész, & Vicsek, 2007) which explains if the group of people know each other, egocentric network (Hollstein, Töpfer, & Pfeffer, 2020) and affiliation network (Lattanzi & Sivakumar, 2009) which demonstrates the interpersonal connection, network of information such as the World Wide Web (Liang, Chan, Zainuddin, & Shah, 2019) and the citation network (Gustafsson, Hancock, & Côté, 2014), the biological network (Girvan & Newman, 2002) such as biochemical network (Barkai & Leibler, 1997) that focuses on the interaction between molecular components, neural network (Vogels, Rajan, & Abbott, 2005) and ecological network (Fath, Scharler, Ulanowicz, & Hannon, 2007) that studies the environmental dynamic, the technological networks such as the Internet, the power grid (Arianos, Bompard, Carbone, & Xue, 2009) that visualizes the whole power grid system, the delivery and distribution network and the transportation network. The pattern of connection in these networks and their structure will help to better understand the corresponding systems.

## 2 Type of Transportation Network

There are a variety of transportation network corresponding to the modes of transportation which are land, air and sea transportation. Transportation network structure differs according to the type of transportation we are dealing with.

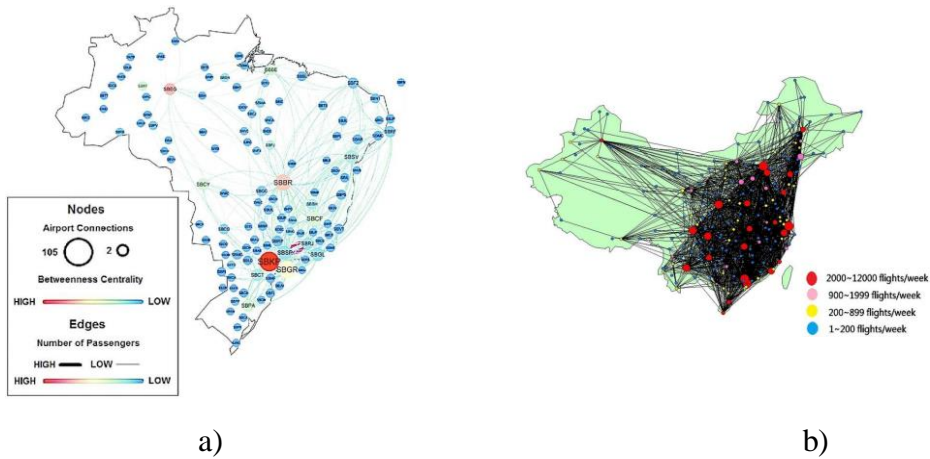

a)                                                                     b)

**Fig. 1**: Examples of transportation network in (a) Brazil by Couto et al. (2015) and (b) China by W.-B. Du et al. (2016)

For sea transportation, the main type of transportation is one that involves freight transportation which is the physical process of transporting commodities and merchandise goods and cargo. Researchers have done massive work on delivery and cargo transportation (Bombelli, Santos, & Tavasszy, 2020). One of the recent works by Hu and Zhu (2009) found that the weighted cluster coefficient and weighted average nearest neighbour degree can help to understand the hierarchy structure and rich-club phenomenon in the network which is influenced by the ship companies' tendency to reduce cost of transportation. Global cargo ship movements on the other hand can help to increase knowledge on the global trade that often act as an economy booster for some countries. From Kaluza, K¨olzsch, Gastner, and Blasius (2010) study, they found that the global cargo shipping network exhibits small world behaviour and is coherent to previous studies on the transportation network based on complex network approach.

For air transportation, Chi et al. (2003) analyse the United States (US) Flight network and concluded that the US flight network exhibits small world behaviour indicated by the high clustering coefficient and small diameter. Moand Wang (2011) conduct an analysis on the air transportation network in China and also obtained similar result. Meanwhile in another study, Gurtner et al. (2014) conducted a community detection analysis on the European airspace as a multi-scale network which includes navigation points, sectors and airport networks. The understanding on air transportation network also can reveal the information on the number of airlines that accommodates the domestic and international flights that may give an overview on the economic growth of a country including the foreign investment of international airline companies (Ladan, 2012). Some of the examples of air transport network is shown in Figure 1.

For land transportation, J. Zhang, Xu, Hong, Wang, and Fei (2011) studied on the topological characteristic and functional properties of the Shanghai subway network in China. The structural analysis on land transportation network such as subway, train or bus can be highly related to the geographical features of the region. As example, the transportation system is mainly focuses on the main cities where there is more suitable land structure (Angeloudis & Fisk, 2006). There are also other types of land transportation such as train service in which Li and Cai (2007) studied on the scale-free railway in China. Analyzing the transportation that involving fixed tracks by applying complex network theory can provide comprehensive evaluation on the tracks particularly in detecting the problem or failures in the tracks (Derrible & Kennedy, 2010). This information is useful as these kinds of transportation is usually used by many citizens and can affect the productivity of the country.

Since there is an increase in the interest of researchers in transportation network, they have been trying to obtain as much information from the network (Debbage, 1999; Ding et al., 2019; Dong, Li, Xing, Duan, & Wu, 2019; Yannis, Kopsacheili, & Klimis, 2012). Some researchers began to analyse transportation network from multiple representation. One of the most common representations is $l$-space network representation (von Ferber, Holovatch, Holovatch, & Palchykov, 2007). This representation is commonly use to study the route network of a transportation system. Other than that, there is also $c$-space representation where the nodes are the routes in the network and the edges are if the routes have a same stop or station (Von Ferber, Holovatch, Holovatch, & Palchykov, 2009). Next, there are also $r$-space representation. This representation is when the starting stop and the terminal stop of a route as nodes, and an edge links between two nodes if these two stop nodes are two ends of a certain route (Huang, Zang, He, & Wei, 2017). Huang et al. (2017) studied on Beijing's weighted train route network and found that the network exhibit scale-free property. The other common presentation widely used by researchers are the $p$-space representation where the node is the stops or stations and there is an edge between all of the nodes that are in the same routes (Sen et al., 2003). This representation is often utilised to study the transfer property of a network. This representation is also known as the transfer network (Sen et al., 2003).

Some researchers utilised more than one representation of the transportation network to give a more comprehensive analysis of the network. For instance, by analysing network in $l$-space and $p$-space representation, one will be able to identify the station that is the busiest meanwhile obtaining information on transfer stations across the network which is a very crucial aspect to travel planning (von Ferber et al., 2007). Y. Zhang, Zhang, and Qiao (2014) studied on Guangzhou metro networks based on the $l$-space and $p$-space representation stated that significant characteristics such as average shortest distance can be obtained from $l$-space and statistical indicators such as average transfer time can be obtained by analysing the $p$-space representation of the network. Von Ferber et al. (2009) studied on the public transportation of major cities of the world by making comparative analysis among all 4 spaces and found that some networks characteristics are more apparent in certain representation compared to the others such as the average shortest distance in $l$-space and average transfer in $p$-space.

A small example of the network is demonstrated in Figure 2. In this example, 3 public transportation network (PTN) routes namely R1 (black line), R2 (blue line) and R3 (green line) as shown below:

**R1** : $A \rightarrow D \rightarrow E \rightarrow C \rightarrow G$

**R2** : $B \rightarrow D \rightarrow E \rightarrow G \rightarrow A$

**R3** : $B \rightarrow C \rightarrow D \rightarrow E \rightarrow F$

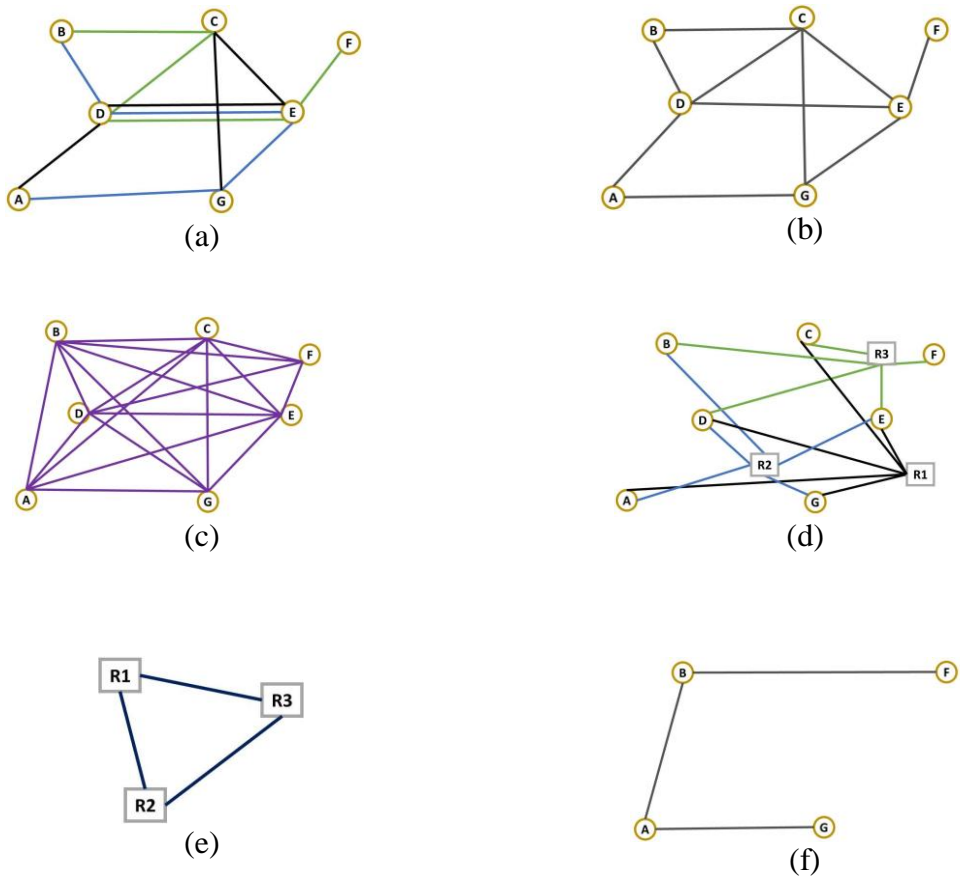

**Fig. 2**: Examples of a) PTN in 5 different spaces namely b) $l$-space, c) $p$-space, d) $b$-space, e) $c$-space and f) $r$-space

## 3 Characteristics of Transportation Network

In transportation network analysis, there are basic measurements that are crucial in studying the network. These basic network characteristics are the nodes, edges, average path length, clustering coefficient, degree and degree distributions (Gattuso & Miriello, 2005). These characteristics can usually help to identify whether the transportation network is scale-free or smallworld network.

Evidently, there are few types of transportation network based on the mode of transportation with different node components. For instance, for the flight route network, the node is the airport (DeLaurentis, Han, & Kotegawa, 2008) while for the airline network, the node is the series number of an airplane since each plane is designated to a different route (Colizza, Barrat, Barthélemy, & Vespignani, 2006). For the maritime network, each node is the ports and links are container liners connecting the ports (Hu & Zhu, 2009). In land transportation, there are different types of public transport namely, bus, train and subway (J. Feng, Li, Mao, Xu, & Bai, 2017). The corresponding networks usually represents their stops or stations as nodes and if the bus, train or subway train went from one route to the other as the edges.

### 3.1 Vertex Degree

Vertex degree or node degree is one of the most important characteristics of the transportation network. There are few types of transportation network representation in which the node for each type of transportation network has a different definition. In $l$-space representation of transportation network or also known as route network, the node degree is the number of routes passing through the particular node which are the stations. In $p$-space representation of transportation network or also known as transfer network, the node degree shows the number of stations that is reachable by passenger without making any transfer into another route (Shi, Lu, Nie, & Yoshitsugu, 2009).

### 3.2 Average Path Length

Average path length is the mean of the shortest path between two nodes also known as mean distance. It is defined as

$$L = \frac{1}{N(N-1)} \sum_{i \neq j} d_{ij} \qquad (1)$$

where $N$ is the number of nodes and $d_{ij}$ is referring to the distance between node $i$ and $j$.

If two nodes are disconnected, meaning there is no path between them, then the path length between them is infinite. If there are two disconnected nodes exist in a network, that will also result the average path length in the network becomes infinite. In Z. Xu and Harriss (2008) study, they computed the average path length of 16 United States intercity passenger airline and found that despite the network consists of large number of cities and connection, almost every two cities are connected in two or less path which is a characteristic of a small-world network. The airport network of India, (Bagler, 2008), the Chinese provincial air transportation network, (W.-B. Du, Liang, Hong, & Lordan, 2017), the urban public transportation systems of 5 Hungarian cities (H´aznagy, Fi, London, & N´emeth, 2015), Beijing bus network (H. Zhang, Zhao, Gao, & Yao, 2013), the Boston and Vienna railway networks, (Seaton & Hackett, 2004) were also found to show small-world characteristics from the observation on their small average path length.

In transfer network or *p*-space network representation, the path length of individual node plays a huge role in identifying the number of transfers between nodes (De Bona et al., 2016; Von Ferber et al., 2009). The measurement is as follows:

$$Number\ of\ tranfers = Path\ length - 1 \qquad (2)$$

This number of transfers refers to the number of times needed for passengers to change their route, making transfers to another route. This can also be described as the number of times the passengers need to get off at a bus stop.

### 3.3 Clustering Coefficient

Clustering coefficient is the measure of the degree to which nodes in a graph tend to cluster together. The clustering coefficient of an undirected graph is the measure of the number of triangles in the graph. In transportation network, clustering coefficient is utilised to describe the level of node cohesiveness in the network (Jia, Ma, & Hu, 2019; Xiaozhou, Xiaoxiao, & Jiang, 2011). Some of the transportation networks are found to have a rather high clustering coefficient indicating the close nodes are well connected (Cheung & Gunes, 2012; Soh et al., 2010). Some of the transportation networks have a smaller clustering coefficient compared to others (Chatterjee, 2015; Couto et al., 2015; Guida & Maria, 2007) which describe the spareness of the distribution of the node in the network. Clustering coefficient are also one of the characteristics that indicates the behaviour of the network which relates directly to the transportation system. This is apparent in determining the location of any important station (Chatterjee, 2015). In the study of United States' Air Transportation (USTN) it is observed that the clustering coefficient is greater than the Worldwide Air Transport Network (WAN), Air Transport Network of India (ANI) and Air Transport Network of China (ANC) which suggest that some airports in United States is very busy compared to other airports in other country (Cheung & Gunes, 2012). Indian Railway Network (IRN) on the other hand, have a high clustering coefficient which explains the number of railway stations that focuses in one area compared to the others (Sen et al., 2003). The formula for the clustering coefficient for node i, Ci is as follows:

$$C_i = \frac{number\ of\ triangles\ connected\ to\ node\ i}{number\ of\ triples\ centered\ around\ node\ i} \quad (3)$$

The clustering coefficient of the whole graph is the average of local clustering coefficient for each individual node. The calculation is as follows:

$$C = \frac{1}{N}\sum_{i=1}^{N} C_i \quad (4)$$

### 3.4 Degree Distribution

Node degree is the number of edges of a particular node in a network. Degree distribution is the probability distribution of the degrees over the whole network. Newman (2010) stated that the degree distribution is the frequency distribution of nodes' degrees. Degree distribution is also a good indicator to identify the transportation network characteristics such a small-work network structure (Guo et al., 2019; X. Xu, Hu, Liu, & Liu, 2007). In Sienkiewicz and Holyst (2005), they studied on the degree distribution of *p*-space representation of 22 public transportation in Poland. The degree distributions are also analysed in world subway networks (Angeloudis & Fisk, 2006), in Kuala Lumpur's public urban rail transit network, (Ding, Ujang, Hamid, & Wu, 2015) where it is observed that these PTN has heavy-tailed distribution with the largest node degree is 2. This observation suggests that the PTN is sparsely connected (Ding et al., 2015).

### 3.5 Dynamical study of Transportation Network

Other than the analysis of the static properties of the transportation network that are done at a fixed time and network structure, many researchers also show some interest in approaching the dynamical perspective of the transportation system (Gao & Jin, 2012; S. Zhang, Derudder, & Witlox, 2016). The difference between the static and dynamical transportation system is that in dynamical systems the changes of the vehicle movement between points in a particular time frame is considered due to the nature of the transportation system service user that is never fixed at all time (Lordan & Sallan, 2020). The passenger movement or frequency alongside the road capability input is among the important features of dynamical approach in transportation system that can contribute to the information on traffic congestion that eventually will affect the network performance (Shen & Gao, 2008).

Another important idea that often being assimilated into dynamical study of transportation network is the Origin-Destination (OD) information (Frederix, Viti, & Tamp`ere, 2013). OD information depicts the movement through geographic area which also usually termed as flow data (Tak, Kim, Byon, Lee, & Yeo, 2018). These flow data could consist of information on the flow of passengers, train and etc. In J. Feng et al. (2017) study, they emphasize on the relationship between both train and passengers flow network wherein provides a thorough perspective on the traffic flow analysis. Studying the train flow system will make it feasible to finally produce a mathematical model that have the ability to estimate the behaviour of a transportation network (But'ko & Prokhorchenko, 2013).

Apart from that, the analysis on hub location is also frequently linked to the assessment of transportation network operation (Aldous, 2008; Alumur & Kara, 2008; Gurtner et al., 2014). In a simpler explanation, hub is the node with a very large degree as compared to the other node. Hub is generally known as a centra collection node in a transportation system or network and mainly used in air transportation network to describe air hubs (Ivy, 1993).

### 3.6 Centralities of Transportation Network

One of the main characteristics that is crucial in transportation network analysis based on complex network approach is centralities. Centrality is the measure of how central a node is based on a few characteristics and measurements. Centrality analysis is one of the prominent analyses that can be observed on a complex network. Centralization analysis is helpful to determine how central any node is in a particular complex network (Tundulyasaree, 2019). In public transportation network especially, centrality is important to identify which node might experience full capacity (Wang, Li, Liu, He, & Wang, 2013). The worldwide air transport network analysis demonstrates that instead of the degree of node, observation on the network centrality is more meaningful as many information on the network global roles can be obtained (Guimera, Mossa, Turtschi, & Amaral, 2005). Meanwhile in China, the centrality assessment on the air transport network shows significant contribution that related to the socioeconomic economic indicators (Wang et al., 2013) such as the development of tourist spot. In the study interregional road of Greece, the centrality measure assists in the representation of Greece interregional commuting system (Tsiotas & Polyzos, 2015).

There are many types of centralities measures that were widely used to characterized complex networks. Some of them are degree centrality, closeness centrality and betweenness centrality. These centralities can also be investigated by plotting their distribution (Chatterjee, Manohar, & Ramadurai, 2016) over all nodes.

### 3.6.1 Degree Centrality

Degree centrality is a centrality measure based on the degree of each node. This centrality measure is one of the commonly used centralities in complex networks to identify the node with high number of degrees which indicating high significance. This centrality is often analysed to study on the connectivity of a network (Wang, Mo, Wang, & Jin, 2011). In Cheng, Lee, Lim, and Zhu (2015) study, they analysed degree centrality of the Singapore subway network and found that the degree centrality is low which is contributed by the characteristic of most nodes that are non-interchange stops have a degree of 1 or 2.

### 3.6.2 Closeness Centrality

Closeness centrality is the measure of to how close is a node to all other nodes based on the shortest path between them. The average value of closeness centrality is regarded as the global accessibility indicator (Tundulyasaree, 2019) This centrality can give information on the accessibility of a network (Wang et al., 2011). The weighted value of the closeness centrality of a network is capable to provide insights on which of the nodes are the critical nodes in the particular network (Z. Du et al., 2020). The closeness centrality value ranges from 0 to 1 indicating value 1 as the node that is directly connected to all the other nodes in the network. The weighted closeness centrality is also regarded as a better way to identify significant nodes compared to the undirected-unweighted network (Z. Du et al., 2020). The closeness centrality also is heavily associated with socio-economic elements (Wang et al., 2011). Closeness centrality measures also helpful in transportation network such as cargo transportation such if there are accidents happen midway, there are the location of which cargo depot to receive assistance such as repair support or technical troubleshooting (Bombelli et al., 2020). In Beijing particularly, the high passenger volume in the air transportation service is explained through the analysis of closeness centrality that focuses on spatial features of the network (Wang et al., 2011).

### 3.6.3 Betweenness Centrality

Betweenness centrality is the measure that are based on if a node being intermediary between others (Ramli, Monterola, Khoon, & Guang, 2014). This centrality shows the significance of nodes as a transfer point between pairs of nodes which later can be used as a transit stops for transportation network especially metro networks (Derrible, 2012; Ramli et al., 2014). For air transportation network, betweenness centrality can be a great indicator to identify which airports could possibly become hub airports (Song & Yeo, 2017). Other than that, betweenness centrality can be used to analyse the resilience of the network by removing the nodes with high betweenness in the network (Wang et al., 2013).

In a route network the node with the highest degree centrality can be analysed as the stops that are tend to be busier than the other stops. The relationship between the degree of a node and its centrality is often studied by researchers to identify the stops that are tend to be busier as compares to others (Cheng et al., 2015). Guimera et al. (2005) stated that there are some nodes that have a small degree while having large centrality. However, they regarded these nodes as oddities as the previous research such as the Internet does not have such characteristics.

## 4 Resilience, Robustness and Vulnerability

One of the important aspects of studying the PTN is to evaluate the PTN performance through network indicators. These network indicators are resilience, robustness and vulnerability.

## 4.1 Resilience

Resilience of a system is connected to the ability to undergo disruptions or failures in the system within an acceptable decrease in performance (Jin, Tang, Sun, & Lee, 2014) which are highly influenced by their structure (Sridhar & Sheth, 2008). Resilience can also be described as how far can the network sustain its connectivity in which when the removal of network component occurred, the presence of a path between pairs of vertices is observed (Tundulyasaree, 2019). One of the examples of the resilience performance of complex network in real-world network is the re-scheduling of journey due to flight failures in airline transportation system (Cardillo et al., 2013). In Berche, Von Ferber, Holovatch, and Holovatch (2009), they studied on 14 major cities public transportation network's reaction towards attacks both on *l*-space and *p*-space network representation. The study shows that the resilience in *l*-space was related to the node degree distribution while in *p*-space the high resilience is related to the small value of average path length. Other than that, network with a large network diameter is bound to be less resilience to failures (X. Zhang, Miller-Hooks, & Denny, 2015). This is because networks with higher network diameter tend to have less redundant connections between pairs of vertices and have sparse links distribution. However, according to X. Zhang et al. (2015) in comparison to diameter, average degree is a better indicator of resiliency as in real-world network that have similar behaviour to small-world and scale-free network model, the nodes have a somewhat drastic differences in the value of node degree. This contributed to some part of the network to have redundant connections while others are more prone to single-link failures.

## 4.2 Robustness

Robustness analysis on the transportation network is based on the performance of the network and the possibility of the occurrence of failures and replacement of paths offered to users (Berche, Ferber, Holovatch, & Holovatch, 2012). Robustness is usually observed after a failure such as removal of a node or edges had occurred (Berche et al., 2012). The change of node size is commonly observed to analyse the network robustness (Yang & An, 2021). Analysis of robustness gained attention of many researchers across the globe as it can provide broad knowledge on the capability of the network.

As example, substitute routes can be suggested during metro network disturbances such as accidents by studying the robustness of metro network (Derrible & Kennedy, 2010). In Madrid, the metro network is deemed as more vulnerable to network changes or disturbances compares to other public transport network after the analysis on the robustness of the network is performed (Frutos Bernal & Martın del Rey, 2019). On the other hand, for the multi-subnet composited complex network consisting bus and subway network, the robustness analysis can be performed by integrating cascading failures in the network before the calculations using the robustness metric (Yang & An, 2021).

## 4.3 Vulnerability

Alongside resilience and robustness, vulnerability is one of the analysis used by the researchers to further expand the perspective on the performance of a network especially public transportation network (Kanwar, Kumar, & Kaushal, 2019). According to Berdica (2002), vulnerability is the susceptibility to incidents that can result in considerable reductions in road network serviceability. The description of vulnerability can be elaborated to two network components, the node and edges (Taylor, 2008). Some researchers also approach transportation vulnerability from accessibility point of view by introducing the accessibility index which includes the traffic flow data (Lu & Peng, 2011). This method later is very useful to study the vulnerability of the road transportation when exposed to disturbances like sea level rise (Lu& Peng, 2011). Other than that, vulnerability can also be studied on a more specific network component such as the edges (Murray-Tuite & Mahmassani, 2004). For transportation network, edges are often used to represent a route between two stops. Vulnerable edges can be identified by ranking the edges according to the availability of substitute paths when network is faced by edge failures such as route closure for land transportation or extreme weather for sea and air transportation.

In addition, the vulnerability of transportation network can also be investigated from the particular line operation (D.J. Sun & Guan, 2016). This can be observed from studying the disruption in every line and eventually analysed the impact on the network vulnerability by taking into account the passengers, average path length and global efficiency of the network before and after disruption circumstances (D.J. Sun & Guan, 2016). One of the approaches of studying the network vulnerability is by assessing the change of relative size of the most connected graph components over a series of subsequent attacks or node removal. If the size of the most connected graph components rapidly decreasing, it is apparent that the network is greatly vulnerable (Xing, Lu, Chen, & Dissanayake, 2017).

## 5 Dynamical Growth and Evolution on Transportation Network

As transportation system is always expanding, it is natural to expect the changes in the structure of the network hence updating the analysis according to the network growth is necessary (P.A. Bonnefoy, 2008). Generally, growth of network is influenced by time as the network progresses and expanding in space (Wang, Jin, Mo, & Wang, 2009). The changes of statistical properties of the network over a period of time such as the connectivity and the network density can explain the evolution of a transportation network (De Montis, Caschili, & Chessa, 2011). In addition, the performance of the transportation network can be identified by examining some parameters such as network density, network centrality, network average distance and network clustering coefficient by making comparative analysis over time and space (Yu, Chen, & Yan, 2019).

From these measurements, some conclusions can be drawn relative to the network size. As example in China, researchers collected information on the air transport network for over the course of 8 decades and come out with a conclusion that the air transport network of China (ATNC) connectivity has become better (Wang, Mo, & Wang, 2014) indicating the ATNC is more accessible despite the increase of the service area. These kinds of observation can only be made if we have information on the transportation network over a period of time. The results from these researches will be an information source for the policy makers in the future to ensure a sustainable development includes the restructuring of transportation network (Kotegawa, DeLaurentis, Noonan, & Post, 2011).

An analysis on the change of the network in a course of a decade for example, can provide valuable information such as recognizing the area that is deemed suitable for public transportation service expansion especially if the network is predicted to enlarge and the public transportation service efficiency is presumed to be improved in the near future (Wei, Mao, Zhao, Zou, & An, 2020). These elements are crucial and often are caused by the increase demand of public transportation network and the accelerated growth of the network (Azzam, Klingauf, & Zock, 2013) caused by the rapid increase in global economy. The surge in demand of the public transportation service will also influence the growth of the PTN (Burghouwt & Hakfoort, 2001). This can be studied by interpreting the scaling mechanism used to cater to the user's demand including expansion in individual stations or even building new complementary stations near the primary station in a particular region (P. Bonnefoy & Hansman, 2007). After the surge in demand is identified, the growth of the PTN is backed by the factors such as capacity, flexibility and safety (X. Sun, Wandelt, & Linke, 2015).

These factors will ensure the expansion of the network is aligned to the need of the users for a better transportation service in terms of performance and accessibility. The structural evolution of the network can be interpreted through a comprehensive analysis of the degree distribution of the network (Mo & Wang, 2011). From the observation of degree distribution, researchers were able to recognize the existence of large cities behaving as the dominant hub as the time progresses (Mo & Wang, 2011). Inarguably, there will be some constraints for the evolution to occur which later affects the statistical characteristics of the network (Colizza et al., 2006). One of the constraints is the spatial constraints that could slow down the evolution of transportation network. The spatial constraints cause the network properties such as the betweenness centrality undergo some unsteadiness when correlated with the node actual positions (Barrat, Barthélemy, & Vespignani, 2005).

Other than studying the network growth and evolution for each country, the global evolution can also be observed by establishing the connections between the main cities of the countries (Wandelt & Sun, 2015). This evolution between countries can often be observed in air transportation network (Wandelt & Sun, 2015). Research by Dai, Derudder, and Liu (2018) study on the evolution of air transportation network in Southeast Asia over 3 decades found that evolution in each country affected the evolution for the whole southeast Asia region with increase of hub cities. While performing an analysis on main cities in a country, the assessment in the evolution of the air transportation network can also help identify each city roles and the contribution to the main air transportation network (Su et al., 2019) such as which city becomes a major hub and the changes of major hubs along the years. The evolution of Chinese main air transport network (CMATN) in a span of 10 years is shown in Figure 3.

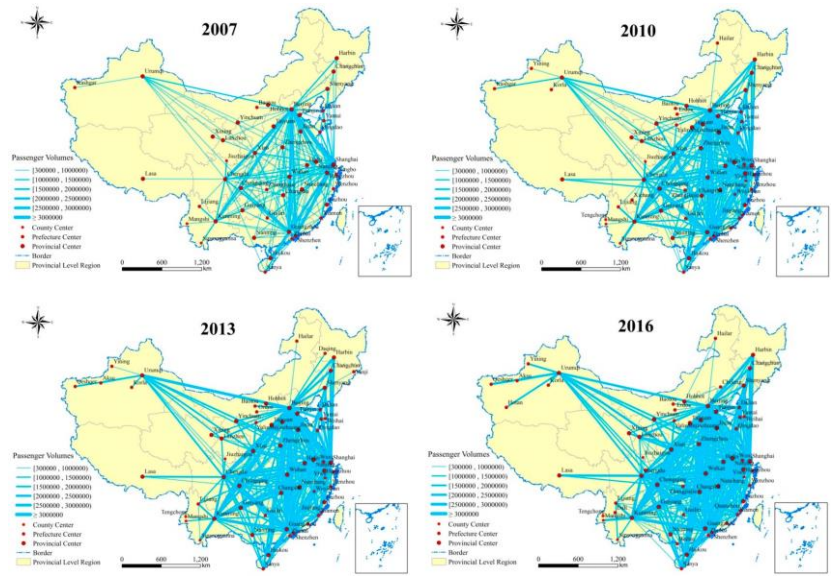

**Fig. 3**: The evolution of Chinese main air transport network (CMATN) in a span of 10 years that demonstrates the increase in passengers volume (Su et al., 2019)

In recent study, S. Feng, Xin, Lv, and Hu (2019) introduced an evolving model of urban rail transit network which is very helpful in urban planning. This model utilised the p-space network structure to better visualize the clustering coefficient and the transfer property between the network components. Meanwhile, Santos and Antunes (2015) introduced an optimization model to increase the efficiency due to the evolution of the airport network in United States. The model considers a variety of factors affecting the evolution includes the capacity, traffic and the utilization cost which covers the congestion tax and airport cost. Interestingly, the connection pattern in the evolution of a transportation network is useful in predicting the spread of global epidemics which was introduced as the global epidemic model by Colizza et al. (2006). This is due to the large-scale properties of the global transportation network and by taking into account the information on the disease infection rate alongside the global traffic flows (Colizza et al., 2006).

## 6 Conclusion

Unquestionably, analysing the transportation network system by adopting the complex network theory is very beneficial to the researchers, service providers and policy makers. Methods of complex network can translate the network structure and network characteristics of the transportation system into real-life especially related to the geographical features of the particular region. Meanwhile, the measurements on the network such as centralities are beneficial to identify significant nodes and transportation network components. The evaluation on the network performance by investigating the network resilience, robustness and vulnerability without a doubt is instrumental in providing a comprehensive analysis on transportation network and its service. Finally, in order to move forward in the fast-pacing world, complex network theory is a very feasible approach in assessing the growth and evolution of the transportation network as it demonstrates the trends over a period of time and increment of space.

**Funding**

The authors wish to acknowledge the financial support provided through the Fundamental Research Grant Scheme (FRGS), Project No. FRGS/1/2020/STG07/UPM/02/2 by Ministry of Higher Education Malaysia.

**Conflict of Interest**

The authors report there are no competing interest to declare.